\documentstyle[twoside,fleqn,espcrc2]{article}

\newcommand{\nc}{\newcommand}
\nc{\be}{\begin{equation}}
\nc{\ee}{\end{equation}}
\nc{\bea}{\begin{eqnarray}}
\nc{\eea}{\end{eqnarray}}

\nc{\eqn}[1]{{(\ref{#1})}}
\nc{\cA}{{\cal A}}
\nc{\cB}{{\cal B}}
\nc{\cC}{{\cal C}}
\nc{\cD}{{\cal D}}
\nc{\cE}{{\cal E}}
\nc{\cF}{{\cal F}}
\nc{\cG}{{\cal G}}
\nc{\cH}{{\cal H}}
\nc{\cI}{{\cal I}}
\nc{\cJ}{{\cal J}}
\nc{\cK}{{\cal K}}
\nc{\cL}{{\cal L}}
\nc{\cM}{{\cal M}}
\nc{\cN}{{\cal N}}
\nc{\cO}{{\cal O}}
\nc{\cP}{{\cal P}}
\nc{\cQ}{{\cal Q}}
\nc{\cR}{{\cal R}}
\nc{\cS}{{\cal S}}
\nc{\cT}{{\cal T}}
\nc{\cU}{{\cal U}}
\nc{\cV}{{\cal V}}
\nc{\cW}{{\cal W}}
\nc{\cX}{{\cal X}}
\nc{\cY}{{\cal Y}}
\nc{\cZ}{{\cal Z}}
\nc{\tr}{{{\rm tr}\,}}

\newcommand{\AmS}{{\protect\the\textfont2
  A\kern-.1667em\lower.5ex\hbox{M}\kern-.125emS}}

\title{Chiral symmetry restoration of QCD and the Gross-Neveu model}

\author{T. Reisz\thanks{Supported by a Heisenberg fellowship,
        e-mail address t.reisz@thphys.uni-heidelberg.de}\address{Institut
        f\"ur Theoretische Physik,\\ 
        Universit\"at Heidelberg, \\
        Philosophenweg 16, \\
        D-69120 Heidelberg, Germany}
}

\begin{document}

\begin{abstract}
Two flavour massless QCD has a second order chiral transition which
has been argued to belong to the universality class
of the $3d$ O(4) spin model.
The arguments have been questioned recently,
and the transition was claimed to be mean field behaved.
We discuss this issue at the example of the $3d$ Gross-Neveu model.
A solution is obtained by applying various well established analytical
methods.
\end{abstract}

\maketitle

%
%
\section{CHIRAL QCD}
Massless QCD has a flavour singlet, chiral symmetry
that is broken spontaneously,
but becomes restored at sufficently high temperature.
For two flavours the symmetry restoring transition is widely
accepted to be of second order.
According to arguments first given by Pisarski and Wilczek,
the universality class belongs to that of the $O(4)$-spin model
in 3d. Beyond others, two main assumptions made are

\begin{itemize}
\item
It is the symmetry pattern that determines the universality class.
\item
Dimensional reduction is complete at the phase transition, that is,
quantum fluctuations along the temperature torus do not 
renormalize the classical 3d IR behaviour.
\end{itemize}

In particular, the second assumption implies that fermions both
do not interact as dynamical degrees of freedom in the IR, and
do not even renormalize the (purely bosonic) effective theory in such
a way that the universality class of the chiral transition changes.

Recently, the $O(4)$ scenario has been questioned, and the
chiral transition was proposed to be mean field behaved.

%
%
\section{THE GROSS-NEVEU MODEL}

The reliability of the arguments against or in favour of the
classical reduction picture is resolved analytically in the
$3d$ Gross-Neveu (GN) model at finite temperature \cite{reisz}.
Here we outline the methods and the results.

The GN model we discuss is a four-point interacting model
of $N$ two-component massless fermions, with partition function
and action
\bea \label{gnaction}
 &&  Z = \int \cD\overline\psi \cD\psi \;
   \exp{\left( - S_{gn}(\overline\psi,\psi) \right)} ,
   \\
 && S_{gn}(\overline\psi, \psi) = \int_x
    \left( \overline\psi \gamma\cdot\partial\psi
  - \frac{\lambda^2}{N} \left(
      \overline\psi \psi \right)^2 \right)(x), \nonumber
\eea
where
$\int_x \equiv \int_0^{T^{-1}} dx_0
\int_{{\bf R}^2} d^2\vec{x}$.
We choose hermitian $\gamma_i$ in the representation 
with $\gamma_0\gamma_1\gamma_2= i {\bf 1}_2$.

The model (\ref{gnaction}) serves as a good example for studying a ``chiral''
transition because it reveals considerable similarities to massless QCD.
In $3d$ the role of chirality is replaced by parity, corresponding to
the transformation
\be \label{gnsymm}
    \psi(x)\to\psi(-x) ,\quad
    \overline\psi(x)\to - \overline\psi(-x).
\ee
The model is invariant under (\ref{gnsymm}). The symmetry would be
broken by a fermionic mass term $m\overline\psi\psi$.
For sufficiently large coupling $\lambda$,
parity symmetry is broken spontaneously at $T=0$,
and it becomes restored at high temperature by a second order
phase transition. The order parameter is given by
\be \label{gncond}
   \frac{1}{N} \; < \overline\psi(x) \psi(x) > ,
\ee
which now represents the parity condensate.

For large $N$ it is convenient to rewrite the model as a Yukawa-type
interacting model, using 
\bea \nonumber
   && \exp{ \left( \frac{\lambda^2}{N}
    \left( \overline\psi \psi \right)^2 \right) }
   \; = \;
   \left( \frac{N}{4\pi\lambda^2} \right)^{\frac{1}{2}} \cdot \\
   && \int_{-\infty}^\infty
   d\sigma \;
   \exp{ \left( - \frac{1}{2} \left( \frac{N}{2\lambda^2} \right)
   \sigma^2
   \pm \sigma \; \overline\psi \psi \right)}.
   \nonumber
\eea
Doing the quadratic fermionic integration, we obtain
\bea \label{gneffs}
  && Z = {\rm c} \; \int \cD\sigma \;
   \exp{ \left( - N \; S_{Y}(\sigma) \right) } , 
  \nonumber \\
  &&  S_{Y}(\sigma) \; = \;
   \int_z \left( \frac{\sigma(z)^2}{4\lambda^2}
   - \tr_s
     \log{K}(z,z)
   \right),
\eea
with $K = \gamma \cdot \partial + \sigma {\bf 1}$,
and the trace is taken over the two spinor indices.
We impose periodic boundary conditions on the auxiliary field $\sigma$
in temperature direction.

In the form (\ref{gneffs}) the GN model is studied by means
of the saddle point expansion for large $N$.
Whereas the expansion in powers of $\lambda$ is non-renormalizable, 
the model becomes renormalizable in the $1/N$-expansion.
Renormalizability provides a large freedom in the choice of the UV
cutoff. For simplicity,
we choose a simple momentum cutoff, which respects parity symmetry.

%
%
\section{PHASE TRANSITION}

The minimizing configurations $\sigma(x)_{min}\equiv\mu$ of the
effective action
(\ref{gneffs}) are obtained as the stable solutions of the gap equation
\be
   \mu \; \left( \frac{1}{4\lambda^2} - J_1(\mu^2,T^{-1}) \right)
   = 0,
\ee
with 
\[
   J_1(x,T^{-1}) = T \sum_{q_0\in\cF}
   \int \frac{ d^2\vec{q} } { (2\pi)^2 } \;
   \frac{1}{q^2 + x},
\]
where $\cF=\{\pi T\nu \vert \nu\in{\bf Z} \;\rm{odd}\}$.
$\mu(T)$ is equal to the parity condensate (\ref{gncond}) to
leading order in $1/N$.

%
%
\subsection{${\bf{N = \infty}}$}

At zero temperature, the broken phase is realized for sufficiently 
large coupling $\lambda$, and we set the $T=0$ mass scale $M$ by
\[
   J_1(M^2, \infty) \equiv \frac{1}{4\lambda^2} < J_1(0, \infty).
\]
As function of $T$, $J_1$ is monotonically decreasing. The parity
condensate $\mu(T)$ approaches zero continuously at
$T_c = M/(2\ln{2})$.
It is straighforward to show that the critical exponents
defined by the parity
condensate and its correlation functions as $T\to T_c$ at $m=0$
(and $m\to 0$ at $T=T_c$)
become $\gamma=1, \nu=1/2, \eta=0, \alpha=0, \beta=1/2$
(and $\delta=3$). 
These are the critical exponents as predicted by mean field 
analysis.

%
%
\subsection{FINITE ${\bf{N}}$}

We solve the question of the universality class by
applying the following two methods.

\begin{itemize}
\item
Dimensional reduction.
For large $N$ the phase transition is described by a $2d$
scalar field theory. There is more than one universality class
with the same symmetry.
By computing the couplings of the effective model,
we identify that part of the phase diagram that corresponds
to the GN model.
\item
Linked cluster expansion.
High temperature series expansions convergent up to the
critical points are applied to the
correlation functions of the effective
model with the prescribed values of the coupling constants.
The high order behaviour
allows for a precise measurement of the critical exponents. 
\end{itemize}

{\bf Dimensional reduction}.
This well established technique applies to local, renormalizable
quantum field theories in $D\geq 3$ dimensions \cite{reisz2}.
For sufficiently high temperature $T\geq T_0$,
and for spatial momenta
$\vec{k}$ sufficiently small w.r.t.~$T$,
the correlation functions are given by
those of a $D-1$ dimensional field theory
at zero temperature, which itself is local and
renormalizable.
\bea
  && <\widetilde\phi(\vec{k}_1);\dots;\widetilde\phi(\vec{k}_n) > \\
  &&= <\widetilde\phi(\vec{k}_1);\dots;\widetilde\phi(\vec{k}_n) >_{eff}
 + O(\frac{\vec{k}_1}{T},\dots,\frac{\vec{k}_n}{T}). \nonumber
\eea
The most general action is obtained by applying a zero-momentum
expansion to the non-static effective action,
\bea
  && S_{eff}^{loc}(\phi_{st}) = \cT_{\phi T^{(3-D)/2},\vec{k}/T}^{n,m,f(n,m)}
  \nonumber \\
  && \lbrack - \ln{\int \cD\phi_{ns} \;
   \exp{(-S(T^{1/2}\phi_{st}+\phi_{ns}) )} }
   \rbrack ,
   \nonumber
\eea
(this provides the loophole in the no-go theorem of
Landsman \cite{landsman}),
where $\cT_{x,y}^{n,m,f}$ denotes the Taylor expansion w.r.t.~$x$
and $y$ of order $n$ and $m$, respectively, constrained by
\[
  f(n,m) \equiv (D-1) - ( \frac{D-3}{2} n + m ) \geq 0.
\]
For the GN model, $T_0(N=\infty)=0$. For sufficiently
large $N$, dimensional reduction works down to the phase transition
temperature $T_c$.
We get with the large $N$ expansion
\bea
 && S_{eff}(\phi) = c(T) \int_{\vec{x}}
  \biggl( \frac{m_0^2}{2}\phi(\vec{x})^2
   +\frac{1}{2}(\partial\phi)(\vec{x})^2 \nonumber \\
 && + T^2 \biggl\lbrack \frac{c_4}{N}\phi(\vec{x})^4
   - \frac{c_6}{N^2}\phi(\vec{x})^6
   + \frac{c_8}{N^3}\phi(\vec{x})^8
  \biggr\rbrack \biggr). \nonumber
\eea
where we have truncated terms of order $N^{-4}$.
All couplings are positive. They are known to leading orders
of the $1/N$ expansion.

{\bf Linked cluster expansion (LCE)}.
According to renormalizability, we are allowed to put the effective model
on the lattice. In a form well known for lattice spin models,
the action becomes in terms of the lattice fields
\bea 
  && \widehat{S}_0(\phi_0) = \sum_{\vec{x}\in\Lambda_2}
   \biggl( - (2\kappa) \sum_{\mu=1,2}
   \phi_0(\vec{x}) \phi_0(\vec{x}+\widehat\mu)
   \nonumber \\ &&
   + \lbrack \phi_0^2
   + \sum_{i=2}^4
     \lambda_i (\phi_0^2-1)^i
     \rbrack(\vec{x}) \biggr). \nonumber
\eea
LCE provides convergent series expansions of connected correlations
and susceptibilities in powers of $\kappa$.
For fixed $\lambda_i$ we have
$\kappa\simeq T^{-1/2}$.
Critical exponents are obtained from
the high order behaviour of the coefficients.
For instance, if 
\[
   \chi_2 = <\widetilde\phi_0(0);\widetilde\phi_0(0) >
    =  \sum_{L\geq 0} a_L \; (2\kappa)^L ,
\]
a critical behaviour as $\chi_2\simeq(\kappa_c-\kappa)^{-\gamma}$
is obtained from
\[
   \frac{a_L}{a_{L-1}} \; = \;
  \frac{1}{2\kappa_c} \left( 1 + \frac{\gamma-1}{L} + o(L^{-1}) \right).
\]
The coefficients are computed to the 20th order.
Some results on the exponents for various $N$ are summarized 
in Table 1.
As a result, the parity restoring transition of the GN model
belongs to the universality class of the $2d$ Ising model.

\begin{table}
\caption{\label{critexp}
The exponents $\gamma$ and $\nu$ of the
GN model.
}
\begin{tabular}{ccc}
\hline
$\quad$ N $\quad$ & $\quad \gamma(N) \quad $ &
$ \quad 2 \nu(N) \quad $   \\
\hline
 46.1  & 1.751(2)  & 2.001(2)     \\
 24.3  & 1.752(3)  & 2.001(4)      \\
 4.68  & 1.749(2)  & 1.999(2)       \\ \hline
\end{tabular}
\end{table}

\section{SUMMARY}

For the Gross-Neveu model with large number $N$ of flavours,
dimensional reduction works down to the phase transition.
The effective $2d$ model has a rich phase structure,
including both Gaussian and Ising model behaviour.
However the values of the coupling constants 
are determined by the reduction
technique. As a result of the computation, the parity transition 
of the Gross-Neveu model corresponds to the Ising part of
the phase diagram.

For QCD we know that dimensional reduction works at least down to
$2T_c$ (\cite{lacock,gupta} and references therein).
Whether it works down to $T_c$ or not is an open question.
There are additional convergence factors $1/T$ compared to the
Gross-Neveu model because of the higher dimension, 
so $N$ is no more required to be large.
However, the renormalized gauge coupling constant becomes large
by approaching $T_c$, so that perturbation theory normally
used in computing the effective action becomes worse.
Other computational techniques are required to realize
the reduction below $2T_c$.

%
%


\begin{thebibliography}{9}
%
\bibitem{reisz} T.~Reisz,
"The Gross-Neveu Model and QCDs Chiral Phase Transition", in
"Field Theoretical Tools for Polymer and Particle Physics",
Lecture Notes in Physics, Vol.~508 (1998) 192
H.~Meyer-Ortmanns and A.~Kl\"umper, eds.
(Springer Verlag, Heidelberg),
and e-print archive hep-lat 9712017.
%
\bibitem{reisz2} T.~Reisz,
Z.\, Phys. {\bf C53}(1992)169
%
\bibitem{landsman}
N.~P.~Landsman,
Nucl.\, Phys. {\bf B322}(1989)498
%
\bibitem{lacock} P.~Lacock and T.~Reisz,
Nucl.\, Phys. (Proc. Suppl.) {\bf B30}(1993)307
%
\bibitem{gupta} S.~Dutta and S.~Gupta,
hep-ph/9806034

%
\end{thebibliography}
\end{document}